# Simi-SFX: A similarity-based conditioning method for controllable sound effect synthesis

**Yunyi Liu, AND Craig Jin**[*]

(yunyi.liu@sydney.edu.au)                                    (craig.jin@sydney.edu.au)

[*]*Computing and Audio Research Laboratory, Department of Eletrical and Information Engineering, The University of Sydney, Sydney, Australia*

Generating sound effects with controllable variations is a challenging task, traditionally addressed using sophisticated physical models that require in-depth knowledge of signal processing parameters and algorithms. In the era of generative and large language models, text has emerged as a common, human-interpretable interface for controlling sound synthesis. However, the discrete and qualitative nature of language tokens makes it difficult to capture subtle timbral variations across different sounds. In this research, we propose a novel similarity-based conditioning method for sound synthesis, leveraging differentiable digital signal processing (DDSP). This approach combines the use of latent space for learning and controlling audio timbre with an intuitive guiding vector, normalized within the range $[0, 1]$, to encode categorical acoustic information. By utilizing pre-trained audio representation models, our method achieves expressive and fine-grained timbre control. To benchmark our approach, we introduce two sound effect datasets–Footstep-set and Impact-set–designed to evaluate both controllability and sound quality. Regression analysis demonstrates that the proposed similarity score effectively controls timbre variations and enables creative applications such as timbre interpolation between discrete classes. Our work provides a robust and versatile framework for sound effect synthesis, bridging the gap between traditional signal processing and modern machine learning techniques.

## 0 INTRODUCTION

Timbre is an audio term used to describe the unique acoustic characteristics of sounds, often referred to as "the color or tone of a sound" [1]. In the realm of sound effects—non-musical and non-speech sounds—timbre serves as a critical audio descriptor, enabling listeners to identify sounds and contributing significantly to the realism and immersive quality of audio experiences [2]. For instance, the sound of footsteps can vary dramatically based on the surface being walked upon. Materials such as wood, gravel, or metal each produce distinct acoustic signatures that reflect their interaction with the environment [3, 4]. The ability to accurately and meaningfully interpolate between these timbres is essential for sound designers aiming to create dynamic and contextually rich soundscapes that mirror varied environmental interactions [5]. This capability is particularly important in industries such as film, gaming, and virtual reality, where nuanced sound design enhances narrative depth and user engagement, ultimately enriching the overall auditory experience [6].

Generating sound effects with controllable variations has been a persistent challenge in audio synthesis [7–9]. Traditional approaches often rely on sophisticated physical models, which demand a deep understanding of signal processing parameters and algorithms. To streamline this process, physically driven neural audio synthesis methods have emerged, leveraging neural networks to reconstruct physical priors from sound, thereby enabling control over the synthesis [10].

While these methods are powerful, they often lack intuitive controls, making them less accessible to non-experts. Additionally, the perceptual subtleties of audio DSP parameters are not always intuitive, posing challenges even for experienced users. The limited availability of diverse, labeled datasets further hinders the development of flexible and broadly applicable sound synthesis systems, underscoring the need for more user-friendly and adaptable approaches.

In the era of neural audio synthesis, generative models are often conditioned on various modalities, such as class labels [11, 12], textual descriptions [13, 14], and visual inputs [15, 16], to guide sound production. While these approaches enable diverse forms of control, they frequently struggle to capture the nuanced timbral differences inherent in distinct environmental contexts. This limitation arises from the reliance on descriptive, hand-crafted labels, which are often insufficient for achieving fine-grained control over sound expression.

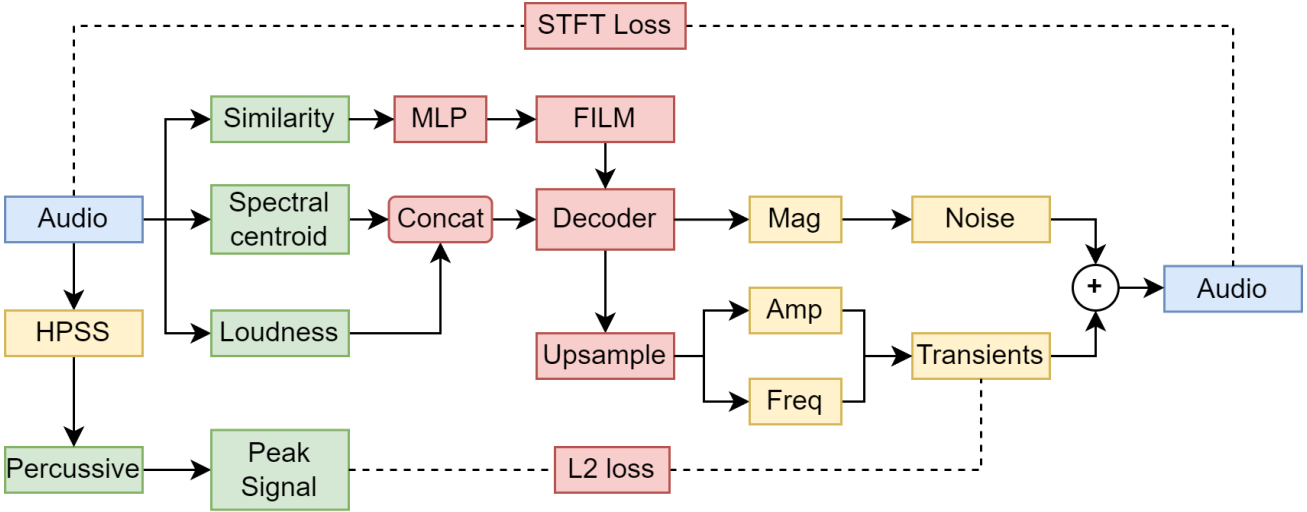

Fig. 1. Overview of our proposed model architecture. Spectral centroid and loudness are extracted explicitly from input audio during the pre-processing stage. Similarity scores are also obtained by measuring the Mahalanobis distance between each audio embedding and group audio embedding of each class. A decoder is conditioned on these three features to generate parameters required for noise and transient synthesizers. The synthesized audio is used to calculate the multi-scale STFT loss for audio reconstruction. Furthermore, to help the transient synthesizer precisely place the transient energies across the time axis, we compute an L2 loss relative to extracted peaks signal from input audio by performing harmonic-percussive source separation.

The Differentiable Digital Signal Processing (DDSP) framework [17] offers a promising alternative by conditioning sound synthesis on detailed features, enabling rapid training and high-quality audio generation. However, despite its strengths, DDSP faces challenges in certain areas, such as morphing between timbres and effectively modeling rigid-body impact sounds—critical components of sound Foley. Addressing these limitations is essential for advancing the flexibility and realism of neural audio synthesis systems.

In this research, we pursue two primary objectives: (1) to generate diverse variations of sounds using a controllable guiding variable that modulates timbre, and (2) to synthesize high-quality inharmonic and impulsive sound effects, such as footsteps and gunshots. While most sound effect datasets are categorized by sound type, we aim to extract and leverage timbral information both within individual categories and across different sound types. This approach captures the similarity of a sound relative to all categories, preserving essential timbre cues that serve as valuable conditioning information.

To ensure high-quality sound synthesis while maintaining computational efficiency for both training and inference, we evaluate our method using a modified DDSP network specifically tailored for synthesizing impulsive broadband sound effects. This framework enables precise control over timbre while meeting the demands of diverse and dynamic soundscapes.

## 1 Related Works

In the domain of neural audio synthesis, the latent space has been extensively explored to control and interpolate the styles of high-level features of the target audio [18–20]. StyleGAN [21] introduces an unsupervised approach that uses intermediate latent vectors to encode different characteristics of the synthesized output, allowing for style mixing. Although originally developed for visual data, this approach can be adapted to sound synthesis to capture and combine diverse stylistic attributes [22]. However, it remains unclear which specific high-level feature is controlled by each sub-latent component of the latent vector.

Given the complexity and intricacy of the latent space, additional effort is often required to establish meaningful control parameters that correspond to specific acoustic attributes. This can be achieved through techniques such as perceptual timbre remapping [23, 24] or adversarial fine-tuning [25]. Audio datasets, particularly in the sound effects domain, often lack detailed descriptive annotations. To address this, many research efforts have focused on extracting meaningful timbral information from audio samples using an analysis-synthesis approach. A straightforward method for achieving a degree of audio control involves using pre-processed audio features, such as pitch [26], loudness [17, 27], or spectral centroid [28], to guide synthesis. These models are typically trained in a supervised learning manner, which often reduces the requirement for extensive datasets.

However, the complexity and diversity of sound effects pose significant challenges. Many variations in sound effects cannot be effectively controlled using only DSP-extracted audio features. For instance, it is exceedingly difficult to transform the timbre of a footstep on a concrete floor into a footstep on snow or sand using solely DSP-based audio descriptors. While class-conditional sound synthesis methods [11, 12, 29] can capture subtle differences between categories, such as different types of footsteps, their capacity for generating new timbres is limited by the discrete nature of one-hot vector representations commonly used in class-conditional models.

To better leverage the information available in sound datasets, Nistal et al. [30] proposed a knowledge distillation approach that enables semantic control over sound categories. This method uses 'soft labels' derived from a large audio tagging system to provide richer conditioning information. Furthermore, Liu et al. [31] introduced an implicit conditioning method that employs continuous vectors sampled from a distribution to represent and condition the generator on each class, enabling smooth timbre interpolation across audio categories. However, since the conditioning vectors are sampled from a standard Gaussian distribution, they are unbounded, which can make them challenging to control effectively.

Unlike pure generative models, differentiable digital signal processing (DDSP) [17] integrates DSP components directly into an autoencoding network. By doing so, it leverages the structured priors provided by DSP principles, which streamline the training process, while also harnessing the neural network's ability to accurately fit the timbre of target sounds. However, the original DDSP employs a harmonic-plus-noise synthesizer, which is primarily designed for harmonic signals and thus inadequate for impulsive sounds, as shown in [27]. For many sound effects that include narrow-band components or transients, the original time-varying FIR filter struggles with the trade-off in time-frequency resolution.

Consequently, several approaches have been proposed to extend DDSP for modeling inharmonic and impulsive sound effects. Lundberg [32] explored vehicle engine noise modeling by integrating a transient synthesizer [33] into DDSP. Liu et al. [27] expanded this idea to model rigid-body impact sounds such as gunshots and footsteps. Serrano [34] focused specifically on modeling footsteps using the original time-varying FIR filter, conditioned on a Ground Reaction Force (GRF) curve. Meanwhile, Diaz et al. [35] developed a differentiable IIR filter bank for modeling rigid-body sounds conditioned on the shape and material of objects. For a more generalized approach, Barahona-Rios [28] introduced a series of multi-rate filterbanks to synthesize source-agnostic sounds with fine-grained control.

## 2 Methods

### 2.1 Similarity Score

Our method for conditioning the generative model draws inspiration from statistical evaluation metrics used in deep generative models. For instance, Frechet Audio Distance (FAD) [36] employs large pre-trained audio representation models, such as VGGish [37] and PANNs [38], to extract meaningful timbre information from both the reference audio dataset and the generated audio. FAD mea-sures the Wasserstein-2 distance between two multivariate Gaussians—one representing real audio embeddings and the other representing generated audio embeddings.

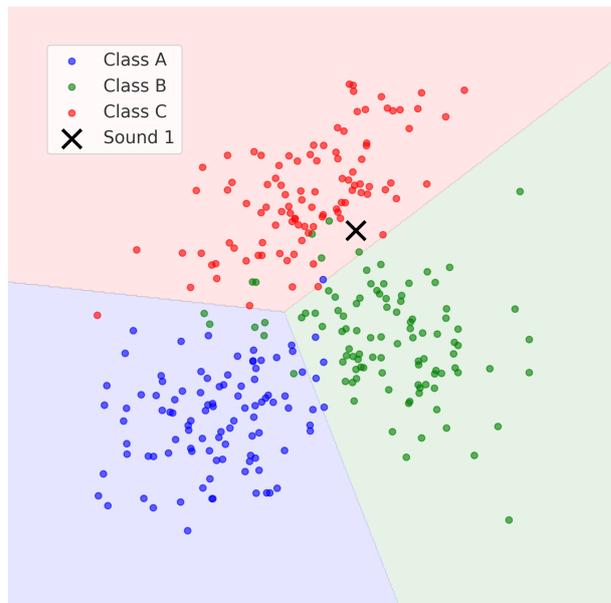

Fig. 2. A tSNE visualization of a three-class embedding clusters. Notice that depending on the distance from each cluster's centroid, a sound might possess the sound characteristics of multiple classes. The closer the sound is from a centroid of a class, the higher the chance that it possess the feature associated with that class.

To illustrate, consider a footstep dataset with three categories, each exhibiting a unique timbre, as shown in Figure 2. For each sound category, we compute embeddings for all audio samples, obtaining the mean and covariance for the category. To extract these embeddings, we utilize the Contrastive Language-Audio Pretraining (CLAP) model [39], which was trained on the extensive LAION-630k dataset containing natural sound effects.

Given an audio signal, such as "Sound 1" in Figure 2, we can quantify its statistical similarity relative to each sound category. Inspired by anomaly detection techniques, where similarity scores are computed to identify outliers, we calculate distances using the Mahalanobis distance (MD) [40]. The Mahalanobis distance is a multivariate measure that assesses the distance between a point and a distribution. In our context, for an audio signal $x$ and a group of audio samples characterized by mean $\mu$ and covariance matrix $\Sigma$, the Mahalanobis Distance (MD) is defined as:

$$MD(x) = \sqrt{(x - \mu)^T \Sigma^{-1} (x - \mu)}, \tag{1}$$

Since the calculation of the MD requires the covariance matrix, $\Sigma$, to be invertible, we regularize $\Sigma$ by adding a small value, a process known as diagonal loading: $\Sigma \leftarrow \Sigma + \varepsilon I$, where $I$ is the identity matrix of the same dimension as $\Sigma$. This ensures the determinant is non-zero. The computed MD can be interpreted as a measure of similar-ity to a class (e.g., how similar a gunshot recording is to a balloon explosion).

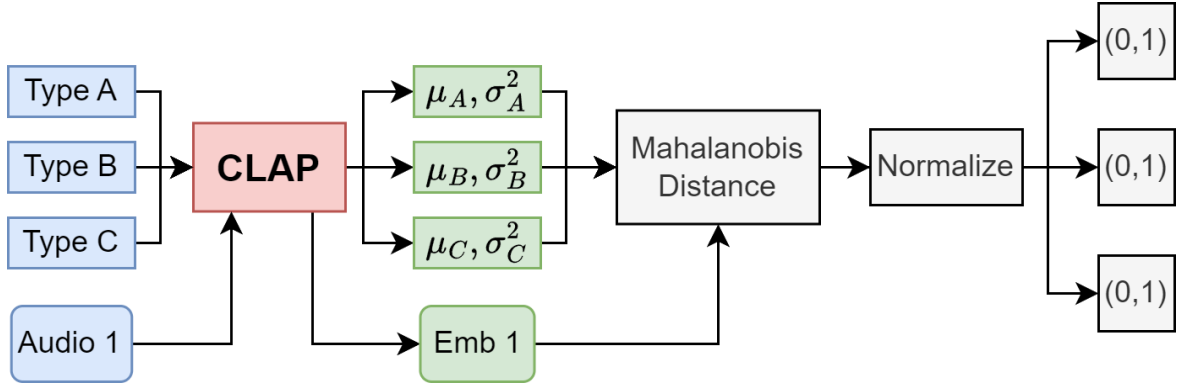

Fig. 3. The process of similarity score extraction. We use Constrastive Language Audio Pretraining (CLAP) to extract embeddings of each sound. The embedding group of audio class can be treated as a cluster of sounds with certain characteristics. For any audio, we can measure the Mahalanois distance relative to each class. The returned score is normalized across each channel to range [0,1] for easy control.

For a dataset with three categories, the MD for a given sound can be computed relative to all three categories, resulting in a three-dimensional vector, as illustrated in 3. However, due to variations in intra-class and inter-class variances, the MD values can exhibit significant range disparities. To address this, we normalize the MD for each channel using min-max normalization, yielding the normalized Mahalanobis distance ($MD_N$):

$$MD_N = \frac{MD - MD_{\min}}{MD_{\max} - MD_{\min}}. \qquad (2)$$

The normalized Mahalanobis Distance falls within the range [0, 1]. A lower MD indicates that the point is closer to the distribution center of the corresponding class, signifying a stronger presence of the timbre associated with that sound category, and vice versa. Using this approach, we derived a compact vector that encodes timbre information, which can be intuitively controlled via a fader or knob [41]. When applied to a generative model, this enables the creation of timbral variations.

### 2.2 Design of the transient model

Our transient modeling approach is founded on a simple yet effective principle: the inverse discrete cosine transform (iDCT) of a sinusoidal signal corresponds to an exponentially decaying impulse, as introduced in [33]. This technique has been successfully applied to DDSP in combination with harmonic-plus-noise synthesis to model various sounds, such as footsteps [27], vehicle engines [32], and piano sounds [42]. Our approach builds upon DDSP-SFX [27], where the decoder outputs frequency and amplitude values to synthesize sinusoids, which are then converted to impulses in the time domain using iDCT. Since multiple transient signals may be required within a single frame of sound, we synthesize $f/2$ sinusoids per frame, where, $f$ denotes the frame length and acts as a sub-sampling rate. The synthesized sinusoids are then modulated by an amplitude vector, $A$, which is learned from the decoder. As the decoder operates at a lower sampling rate, $A$ is upsampled to the target sampling rate $Fs$ to match the length of the entire signal. This upsampling is achieved using a combination of transposed 1-D convolution and ReLU activation. The complete transient model can be formally expressed as:

$$x[n] = \sum_{n=0}^{N-1} \sum_{k=1}^{K} A_{n,k} \, \text{IDCT} \left( \sin \left( 2\pi F_{n,k} \frac{t}{f} \right) \right), \qquad (3)$$

where:

- $K$ is the number of sinusoids synthesized per frame and $K = f/2$ in our case.
- $A_{n,k}$ is the amplitude of the $k$-th sinusoid in the $n$-th frame.
- $F_{n,k}$ is the frequency of the $k$-th sinusoid in the $n$-th frame.
- IDCT is the inverse discrete cosine transform.
- $t$ is the time index.
- $f$ is the frame length.

### 2.3 Loss function for the transient model

As the FIR-filtered noise synthesizer operates at a lower sampling rate, it may inadequately capture the details of transient energies. NoisebandNet [28] addresses the tradeoff between frequency and time resolution by designing a large-scale filterbank, albeit at the expense of increased computational cost. However, equidistant placement of transients along the time axis, as proposed in [32, 33], does not fully resolve the issue of missing transient energies.

To better guide transient synthesis, we propose incorporating a dedicated loss function. First, we separate the percussive components of the training audio using harmonic-percussive source separation (HPSS) [43], with a kernel size of 31 and a margin of $(1, 3)$. Next, local peak estimation is performed to identify spectral peaks along the time axis, which are then converted into one-hot vectors, with non-peak regions set to 0. To ensure that synthesized impulses are placed only at these peak locations, we create a sparse waveform $X_s$ by multiplying the one-hot peak vec-tors with the original waveform. Finally, an $L_2$ loss is computed directly between the transient signal and the sparse peak waveform, providing a more targeted approach for transient synthesis.

## 2.4 Architecture

In Figure 1, we present our model architecture, which builds upon DDSP. Similar to NoisebandNet [28], we extract spectral similarity and loudness features from the input audio, sampled at $F_s = 44100$ Hz, using a window size of $W = 256$. These features are concatenated and passed into the decoder to generate parameters for the noise and transient synthesizers. Additionally, we compute a similarity vector by calculating the MD relative to each class, as illustrated in Figure 3. The similarity vector is an $n$-channel vector where $n$ represents the total number of classes. This vector is first smoothed using a Multi-Layer Perceptron (MLP) layer. Subsequently, we employ Feature-wise Linear Modulation (FiLM) [44] to project the similarity vector onto the decoder with a sequence length matching that of the input features. We utilize the same decoder architecture as DDSP [17], which is responsible for generating the magnitude frequency response for the noise synthesizer, as well as the amplitude and frequency parameters for the transient synthesizer. The outputs of the noise and transient synthesizers are combined and subsequently processed using the same reverb algorithm as DDSP [17].

## 3 Experiments

### 3.1 Dataset

Table 1. Number of Sounds in Footstep and Impact Categories

| Footsteps | | Impacts | |
|---|---|---|---|
| Category | Amount | Category | Amount |
| All | 2127 | All | 1666 |
| Board | 83 | Smash | 153 |
| Concrete | 278 | Gunshots | 280 |
| Gravel | 292 | Knocks | 321 |
| Leaves | 378 | Explosion | 169 |
| Snow | 253 | Footsteps | 528 |
| Squeaky | 65 | Bounce | 155 |
| Stairs | 207 | Punch | 60 |
| Tile | 201 | | |
| Wood | 370 | | |

In this research, we focus primarily on the generation of rigid-body impact sounds. Most existing sound effect datasets categorize sounds based on broad classifications (e.g., footsteps, rain, birds), with limited detailed separation within categories (e.g., footsteps recorded on different materials). To address this, we sourced 2,127 footstep sounds from Freesound [45] and categorized them into nine groups based on the material of the contacting surface, as detailed in Table 3.1. The timbral differences across these categories provide a suitable framework for testing the effectiveness of our proposed conditioning variable in controlling the timbre of synthesized sounds. In addition to the Footstep-set, we collected another 1,666 impact sounds commonly used in games from Freesound. This broader classification allows us to evaluate the effectiveness of our conditioning approach under greater variance across class embeddings. All sounds were sampled at 44.1 kHz with 16-bit depth and a duration of 4 seconds. The datasets can be found at: https://zenodo.org/records/14286414

### 3.2 Training

To set up the training, we first extract spectral centroid and loudness features from each audio sample, using an FFT size of 256 and 50% overlap. This process results in feature vectors with a sequence length of 690. For the decoder, we use a hidden size of 512 across all experiments. The noise synthesizer is configured with 100 frequency bands. As detailed in Section 3, we use 128 frequencies for sinusoidal modeling, which corresponds to half the frame size. To prevent aliasing, the frequencies are clipped to the range $(0, 128)$. For the multi-scale STFT loss [46], we compute spectrograms using FFT sizes of 2048, 1024, etc. down to 16, dividing by a factor of two. The model is trained for 5,000 epochs on an NVIDIA RTX 4070 Ti GPU, with a batch size of 16. We use the ADAM optimizer [47] with an initial learning rate of of $1 \times 10^{-4}$, which decays to $1 \times 10^{-5}$ after 80% of the training epochs. Training takes approximately 17 hours for the Footstep-set and 13 hours for the Impact-set.

### 3.3 Fine-tuning based on similarity

Since the extracted similarity score is normalized to the range $[0, 1]$, its distribution is not constrained to follow any specific form, unlike in ICGAN [31]. To better understand its probability density, we perform Kernel Density Estimation (KDE) as follows:

$$\hat{f}(x) = \frac{1}{nh} \sum_{i=1}^{n} K\left(\frac{x - x_i}{h}\right) \quad (4)$$

where $\hat{f}(x)$ is the estimated density function at point $x$, $n$ is the number of data points, $x_i$ are the data points, $K(\cdot)$ is a Gaussian kernel function, $K(u) = \frac{1}{\sqrt{2\pi}} e^{-\frac{u^2}{2}}$, and $h$ is the bandwidth, a positive scale factor that affects the smoothness of the density estimation. In Figure 4, we present the KDE results for each category in our footsteps dataset. The plots reveal significant variability in the probability density of similarity scores across classes. For instance, certain ranges of similarity scores (e.g., 0.8-1.0 for concrete) are underutilized during training compared to others (e.g., 0.3-0.5 for concrete). This imbalance can hinder the model's ability to generate diverse timbre variations. To address this, after the model achieves satisfactory synthesis performance, we fine-tune the similarity conditioning layers to account for the underutilized ranges. Instead of using the similarity scores extracted from the dataset, we input pseudo-scores sampled uniformly from the range $[0, 1]$. This ensures that all score regions are captured and utilized for guiding sound synthesis. As illustrated in Fig-ure 5, during fine-tuning, we freeze the decoder layers and update only the similarity conditioning layers. Since the objective is no longer to improve synthesis quality, we omit the reconstruction multi-scale STFT loss. Instead, we treat fine-tuning as a regression task, minimizing the error between the ground-truth similarity score (sampled from the uniform distribution) and the score measured from the generated audio. The loss function combines $L_1$ and $L_2$ losses and is defined as follows:

$$\mathscr{L} = \frac{1}{n}\sum_{i=1}^{n}(s_i - \hat{s}_i)^2 + \frac{1}{n}\sum_{i=1}^{n}|s_i - \hat{s}_i| \tag{5}$$

where $s_i$ represents the ground-truth similarity score, $\hat{s}_i$ is the predicted score and n is the batch size. Similar to the training process, we employ a batch size of 16 and a learning rate of $1 \times 10^{-4}$ with the ADAM optimizer [47]. The fine-tuning is conducted for a total of 10,000 epochs on an NVIDIA 4070 Ti GPU. This process takes approximately 10 hours for the Footstep-set and 8 hours for the Impact-set.

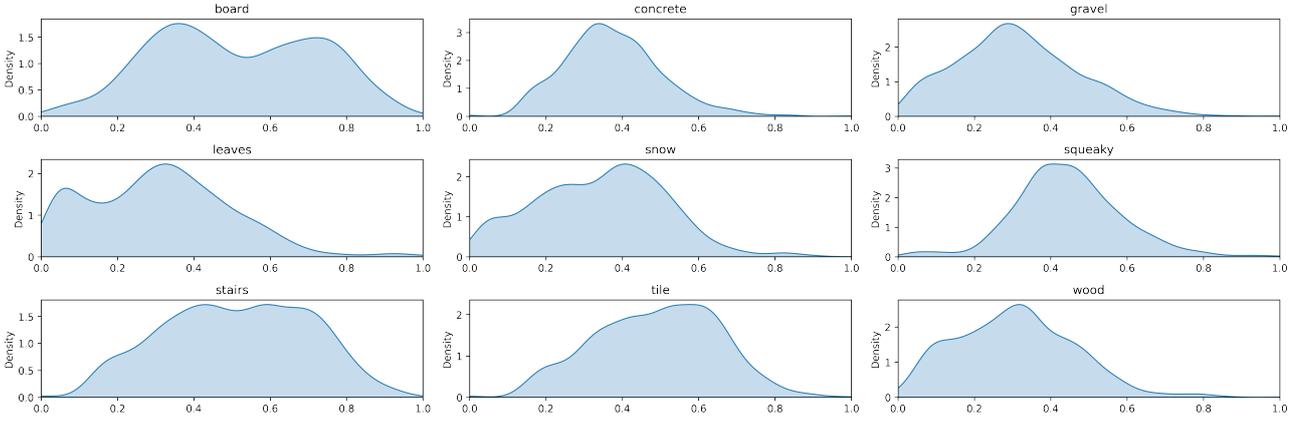

Fig. 4. Kernel Density Estimation of the distribution of similarity scores of Footstep-set. The horizontal axis represents the similarity scores, ranged from [0,1]. The vertical axis shows the estimated density values. We noticed different distributions of similarity scores across different categories.

## 4 Evaluation

### 4.1 Synthesis Performance

We evaluate the synthesis performance of our method against three neural audio synthesis models: the original DDSP (using only subtractive noise synthesis), the Noise-bandNet model, and ICGAN [31].

For the DDSP model, we maintain the same configuration for its noise synthesizer and training settings as in our approach. This includes parameters such as the number of FFTs, multi-scale STFT loss, batch size, and training epochs. The model is also guided by spectral centroid and loudness features, extracted in the same manner as in our method, to inform the decoder during sound synthesis.

For the NoisebandNet model, we implement its filter-bank using 2,048 frequency bands, a hidden size of 128, and a window size of 32. The model is trained separately on our two proposed datasets, using 10,000 epochs, a batch size of 16, and a learning rate of 0.001. The ICGAN model, which operates in the spectral-temporal domain, requires the extraction of Mel-spectrograms from each audio sample. These spectrograms are computed with a time resolution of 690 and 64 frequency bins. The model is then trained for 10,000 epochs with a batch size of 16 and a learning rate of $1e-4$.

To evaluate synthesis performance, we use two objective metrics: Frechet Audio Distance (FAD) [36] and Log Spectral Distance (LSD). FAD is a statistical evaluation metric that measures the overall similarity between two distributions: the reference dataset and the reconstructed dataset. In contrast, LSD provides a pairwise comparison between a reference sound and its corresponding generated sound. To generate the reconstructed dataset, we use the test sets of our datasets as input audio. For each test sample, we extract the required audio features to guide the models and set each model to evaluation mode to synthesize the reconstructed audio. This ensures that every reference sound in the test set is paired with a corresponding generated track from each model. FAD and LSD calculations are then performed using these reference and generated datasets, allowing us to objectively assess the synthesis performance of each model.

### 4.2 Objective measure of controllability

In our study, we employ regression analysis, specifically Ordinary Least Squares (OLS) [48], to quantitatively evaluate the effectiveness of the conditioning method used in audio signal synthesis. Regression analysis is a robust statistical technique for quantifying the relationship between independent variables and a dependent variable. The OLS method, a fundamental and widely used form of regression analysis, estimates the parameters of a linear regression model by minimizing the sum of squared residuals. This approach ensures that the best-fit line closely approximates the true data points. In our analysis, the independent variable, $c$, represents similarity scores interpolated between 0 and 1, indicating the proximity to a specific class $A$ (rang-ing from close to distant). The dependent variable, $y$, is the measured similarity of the synthesized sound relative to each class. Ideally, when $c = 0$, the synthesized sound should exhibit a lower distance to class $A$, indicating high similarity. Conversely, when $c = 1$, the synthesized sound should demonstrate a higher distance, reflecting low similarity. This setup allows us to directly evaluate the model's ability to maintain meaningful conditioning during synthesis.

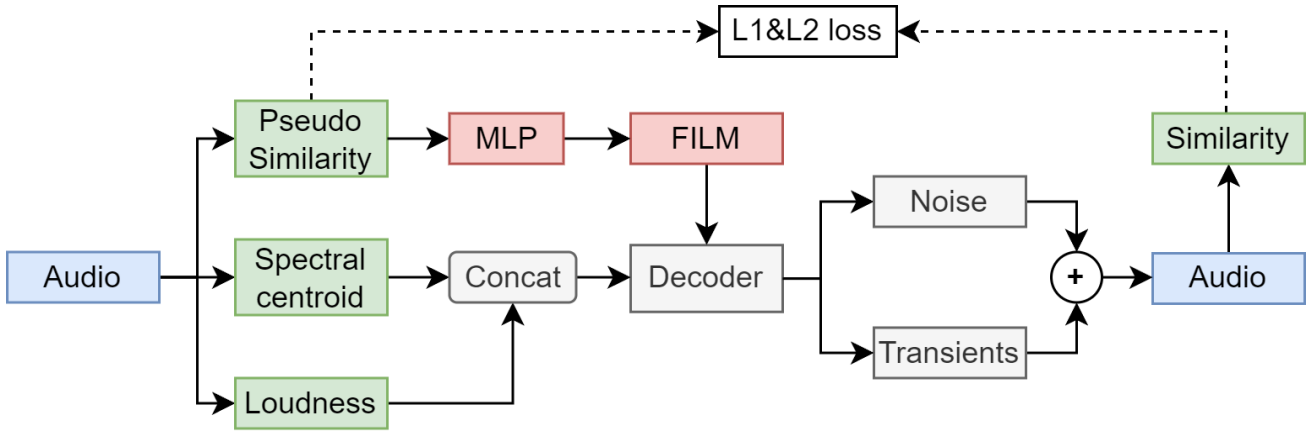

Fig. 5. Fine-tuning process. The weights of the decoder are completely frozen during the fine-tuning stage. For each batch, we sample a vector from a uniform distribution in [0,1] and pass it to the decoder. We then measure the similarity score of the generated audio and evaluate a mixture of L1&L2 loss. In this way, the MLP + FILM layers are trained to match the input similarity scores with generated sounds with diverse timbres.

### 4.3 Regression Analysis Setup

To set up the experiment, we generated variations of each reference sound in our test set by interpolating the similarity score for one channel from 0 to 1 in 100 steps, while keeping all other channels fixed at 1. This resulted in 100 synthesized versions of each sound, all guided by the same reference. To measure the similarity of each output sound relative to each category, we opted to abandon the pre-trained CLAP model for embedding extraction to minimize potential biases. Instead, we used PANNs [38], a model extensively employed in audio tagging and sound event detection. A PANNs model pre-trained on Audioset [49] was fine-tuned to classify the synthesized sounds into their respective categories. After fine-tuning, the model achieved a classification accuracy of 94.7%. Subsequently, we removed the final layer of the model to use the remaining network as an embedding extractor.

Using these embeddings, we calculated the Mahalanobis Distance (MD) of each synthesized sound's embedding relative to the embedding clusters of each class, following a process similar to the one described in Figure 3. For consistency and ease of visualization, the MD values for each channel were normalized to the range [0, 1]. Finally, each synthesized sound's similarity metric was paired with its corresponding guiding vector value, creating a dataset with independent variables (interpolated guiding vectors) and dependent variables (normalized MD values). This dataset forms the basis for evaluating the relationship between the guiding vectors and the resulting synthesized sound similarities.

Using Ordinary Least Squares (OLS) regression, we analyzed the relationship between the guiding vector values–linearly interpolated from 1 (least similar) to 0 (most similar)–and the normalized Mahalanobis distances, which represent the timbral similarity to each target class. OLS was chosen for this analysis because of its ability to provide straightforward parameter estimation by minimizing the sum of squared differences between observed and predicted values in a linear model. Through preliminary analysis, we observed that the relationship between the guiding vector values and the normalized distances is better described by an exponential rather than a linear trend. To accommodate this, we transformed the exponential regression model into a linear form by taking the natural logarithm of both sides. This transformation allows us to perform OLS regression on the modified model:

$$\ln(y) = \ln(a) + b \cdot x \tag{6}$$

where $y$ is the dependent variable (normalized MD), $x$ is the independent variable, $\ln(a)$ is the transformed coefficient to be estimated as the intercept, and $b$ remains the slope of the regression line. The objective of ordinary least squares (OLS) regression is to minimize the sum of squared residuals:

$$\min_{\ln(a),b} \sum_{i=1}^{n} \left( \ln(y_i) - (\ln(a) + b \cdot x_i) \right)^2 \tag{7}$$

where $n$ is the number of interpolation steps. After estimating the parameters, the model for predictions is given by:

$$y_{\text{pred}} = a \cdot e^{b \cdot x_{\text{pred}}} \tag{8}$$

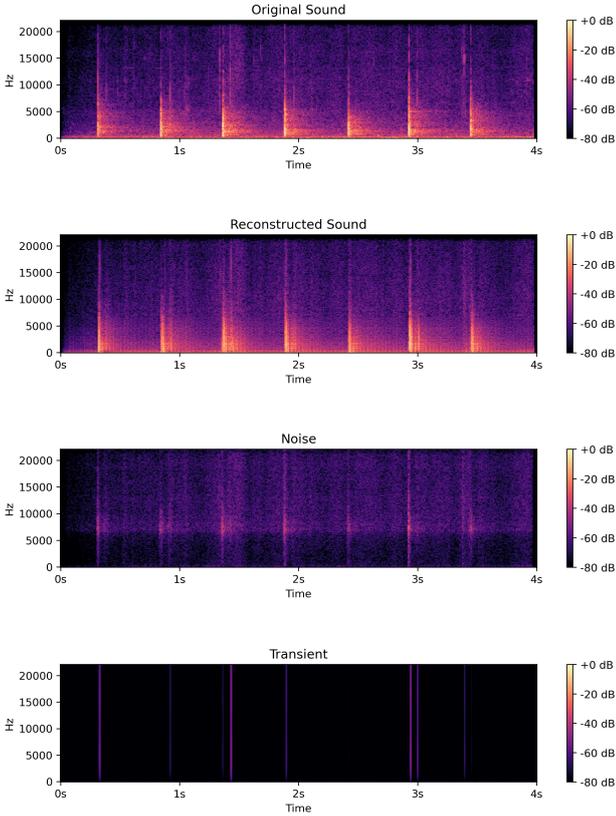

Fig. 6. Sound reconstruction based on a reference track. The spectral centroid, loudness envelope, as well as similarity scores are extracted from the original sound and used for audio reconstruction. The reconstructed, noise, and transient separations are shown from top to bottom.

## 5 Results

### 5.1 Synthesis Performance

An impactful footstep often contains a sudden burst of energy that is challenging to model solely by filtering white noise through a linear time-varying finite impulse response (LTV-FIR) filter. In Figure 6, we present a randomly selected footstep sound (5.1), its reconstructed version using our proposed approach (5.1), the synthesized noise component (5.1), and the synthesized transient component (5.1). The reconstructed sound effectively captures the spectral content of the reference sound.

While the noise component, modeled with the LTV-FIR filter (using 256 FFTs), accurately reproduces the frequency spectrum of the reference audio, it falls short in representing the temporal dynamics characteristic of the original sound. This limitation is evident in the lack of temporal sharpness and impact, which are crucial for delivering the precise auditory cues associated with the source material.

To address these shortcomings, we incorporate transient synthesis (illustrated as vertical impulses with evenly distributed frequencies over a very short duration). This addition enhances the model's capability to reproduce the rapid onset and decay times that contribute to the immediacy and impact of natural sounds. Moreover, by regularizing tran-sient synthesis with an additional $L_2$ loss based on a sparse, peak-detected signal, the decoder learns to emphasize amplitudes specifically at onset locations while attenuating amplitudes elsewhere. This targeted approach improves the realism and perceptual quality of the synthesized sounds.

Table 2. Model Performance Comparison Across Datasets and Measurements.

| Models | Footsteps | Impacts |
|---|---|---|
| **FAD↓** | | |
| **Ours** | $6.0 \pm 1.39$ | $2.06 \pm 0.91$ |
| **DDSP** | $10.0 \pm 2.39$ | $6.07 \pm 3.53$ |
| **Noisebandnet** | $4.06 \pm 1.08$ | $1.89 \pm 1.04$ |
| **ICGAN** | $12.52 \pm 4.81$ | $4.92 \pm 0.84$ |
| **LSD↓** | | |
| **Ours** | $0.14 \pm 0.03$ | $0.11 \pm 0.05$ |
| **DDSP** | $0.22 \pm 0.05$ | $0.30 \pm 0.11$ |
| **Noisebandnet** | $0.13 \pm 0.02$ | $0.11 \pm 0.04$ |
| **ICGAN** | $0.20 \pm 0.06$ | $0.18 \pm 0.08$ |

Since our datasets consist of multiple categories, we computed the mean and standard deviations for the entire dataset to ensure robust evaluation. Both the FAD and LSD results demonstrate that the performance of our model is comparable to that of the NoisebandNet model, which employs a large filterbank noise synthesizer with pre-stored noise bands. Although our architecture uses the same noise synthesizer as the original DDSP, the addition of the transient synthesizer significantly enhances the placement of transient energies at a sub-frame level. This improvement is crucial for reconstructing fine-grained details of impulsive sounds.

The ICGAN model performs similarly to DDSP but falls short of delivering satisfactory quality compared to the other models. We hypothesize that this shortfall may be due to the implicit nature of its conditioning method for encoding class information. This approach introduces stochasticity into the conditioning process, potentially degrading audio fidelity.

Additionally, we observed significant differences in results across the two datasets we evaluated. The Impact-set achieved better results than the Footstep-set, despite identical configurations for both datasets. We hypothesize that this disparity arises from the lower variance in the Impact-set compared to the Footstep-set.

Overall, our approach demonstrates its effectiveness for synthesizing sound effects rich in dynamics and transients. It achieves high-quality audio synthesis without the need to train a large generative model, making it a practical solution for various sound synthesis applications.

### 5.2 Conditioning performance

In Figure 8, we present the plots of measured normalized Mahalanobis Distances (MD) as a function of interpolated similarity scores ranging from 0 to 1. The relationship between the normalized MD of each synthesized sound and its conditioning similarity score closely follows an exponential trend. Ordinary Least Squares (OLS) regression yields a mean $R^2$ value of 0.4774 for the Footstep-set model and 0.6041 for the Impact-set model, indicating a strong correlation between the independent variable $c$ and the dependent variable $y$.

The plots reveal that almost all sound categories exhibit clear separations between the presence of a particular feature (at $c = 0$) and its absence (at $c = 1$). This demonstrates that the model successfully learns to output distinct timbres based on the proposed similarity scores. Additionally, the regression lines indicate a positive correlation between the normalized MD and the input similarity scores. This suggests that the conditioning method effectively encodes timbral information unique to different sounds, independent of other acoustic features such as loudness and spectral centroid, which were held constant during this test.

Notably, a clear separation is observed in the similarity score range of 0.8 to 1.0 across most categories. This implies that values below 0.8 have limited influence on encoding timbre variations, potentially indicating a model bias toward generating outputs with specific modes. To address this limitation, future research could explore strategies to expand the effective range of control. This may involve employing pre-processing techniques to flatten the similarity score distribution or mapping the scores to a more meaningful scale to better utilize the entire range.

## 6 Creative Usage of Timbre Control

In Figure 7, we qualitatively demonstrate how interpolating between two conditioning similarity scores produces distinct timbres. For this example, we randomly selected a sound from our test dataset and extracted its loudness and spectral centroid as input. All channels of the similarity score were fixed at 1, except for the first channel ($C_1$) and the second channel ($C_2$) which correspond to footsteps on metallic boards and footsteps on gravel, respectively. These channels were interpolated from 0 to 1 ($C_1$) and 1 to 0 ($C_2$).

The resulting spectrograms reveal a clear progression in timbre. Initially, the signals exhibit prominent harmonics in the higher frequencies, characteristic of footsteps on metallic boards. As the interpolation progresses, these harmonics gradually transition into noisier signals lacking harmonic structure, which are indicative of footsteps on gravel. For a more detailed visualization and auditory experience, please visit our accompanying website: `https://reinliu.github.io/Simi-SFX/`.

## 7 CONCLUSION

In this research, we proposed a similarity-based conditioning method designed to control timbre variations in sound effects. This method was implemented within a lightweight neural audio synthesis model and evaluated using our custom sound effect dataset, demonstrating its creative potential for sound design. We showed that this conditioning approach enables the interpolation of subtle sound characteristics unique to different sound categories, providing nuanced control over the synthesis process.

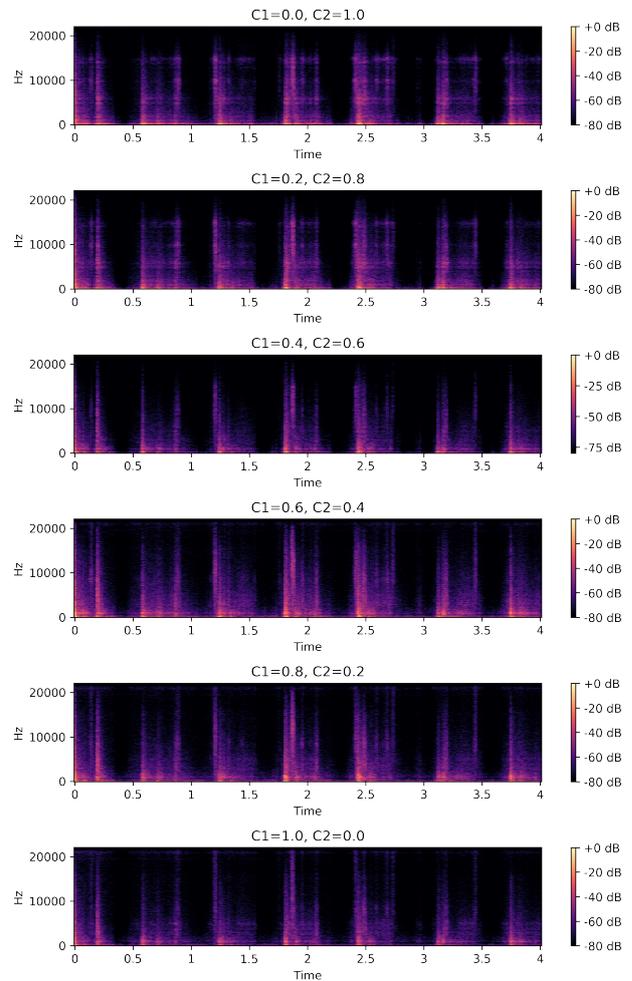

Fig. 7. An example of interpolating between two conditioning similarity scores. We fixed all other channels as 1 for the whole time, and interpolated the first channel $C_1$ (Footstep: Board) from 0 to 1, while the second channel $C_2$ (Footstep: Gravel) from 1 to 0. A lower conditioning score of $C_1$ or $C_2$ indicates a higher similarity with class one or two, respectively.

The synthesis quality achieved by our method is comparable to state-of-the-art data-driven sound effect synthesis algorithms. Furthermore, we conducted regression analysis to evaluate the relationship between the synthesized sounds and the conditioning vectors. The results indicate that our method effectively separates sound characteristics along the extremes of the conditioning vectors, validating its ability to generate distinct timbral variations.

However, we observed that the effective range of control is densely concentrated around similarity scores of 0.8 to 1.0, limiting the method's overall controllability. Future research could focus on improving the distribution of the conditioning similarity scores, ensuring a flatter, more uniform range to enhance the granularity and effectiveness of timbre control.

**APPENDIX**

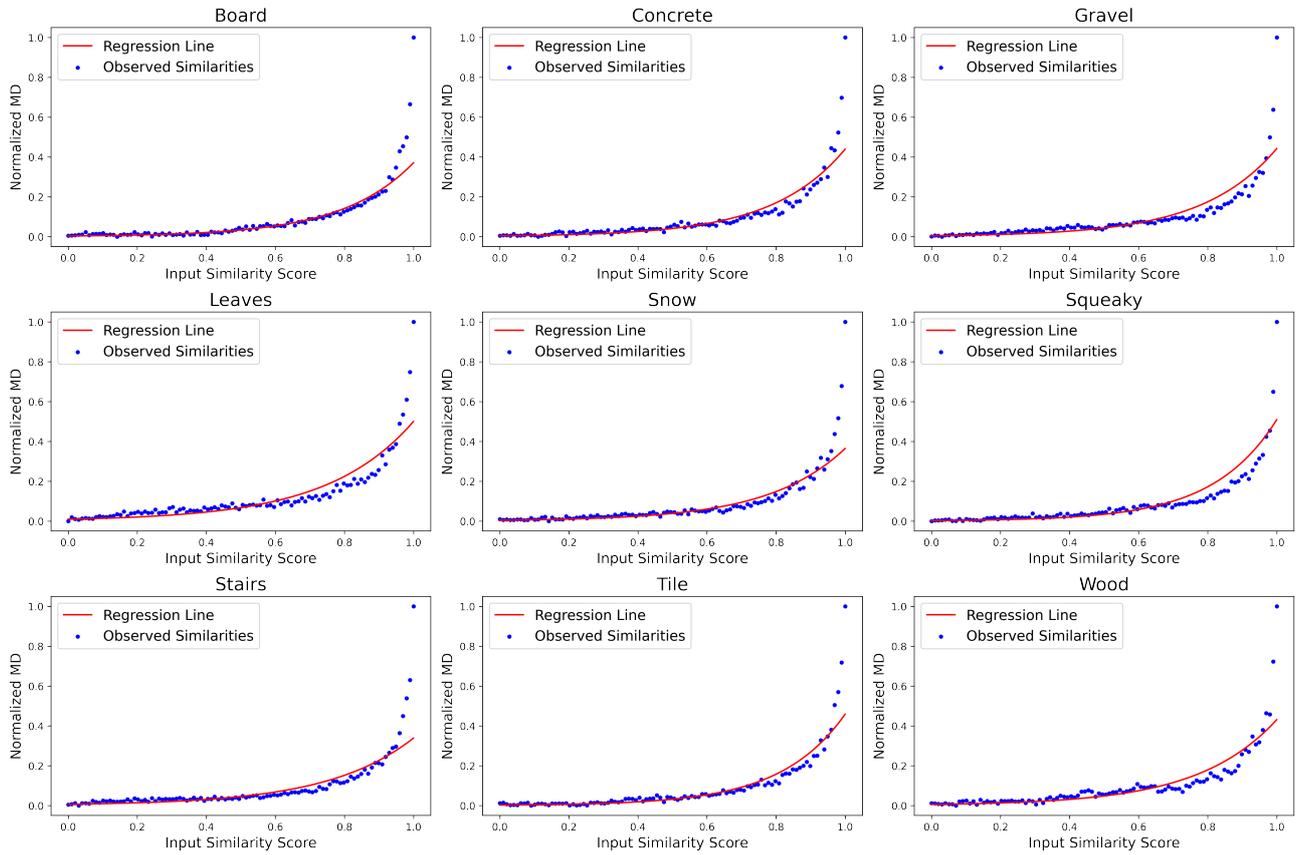

Fig. 8. Regression Analysis for our model trained on Footstep-set.

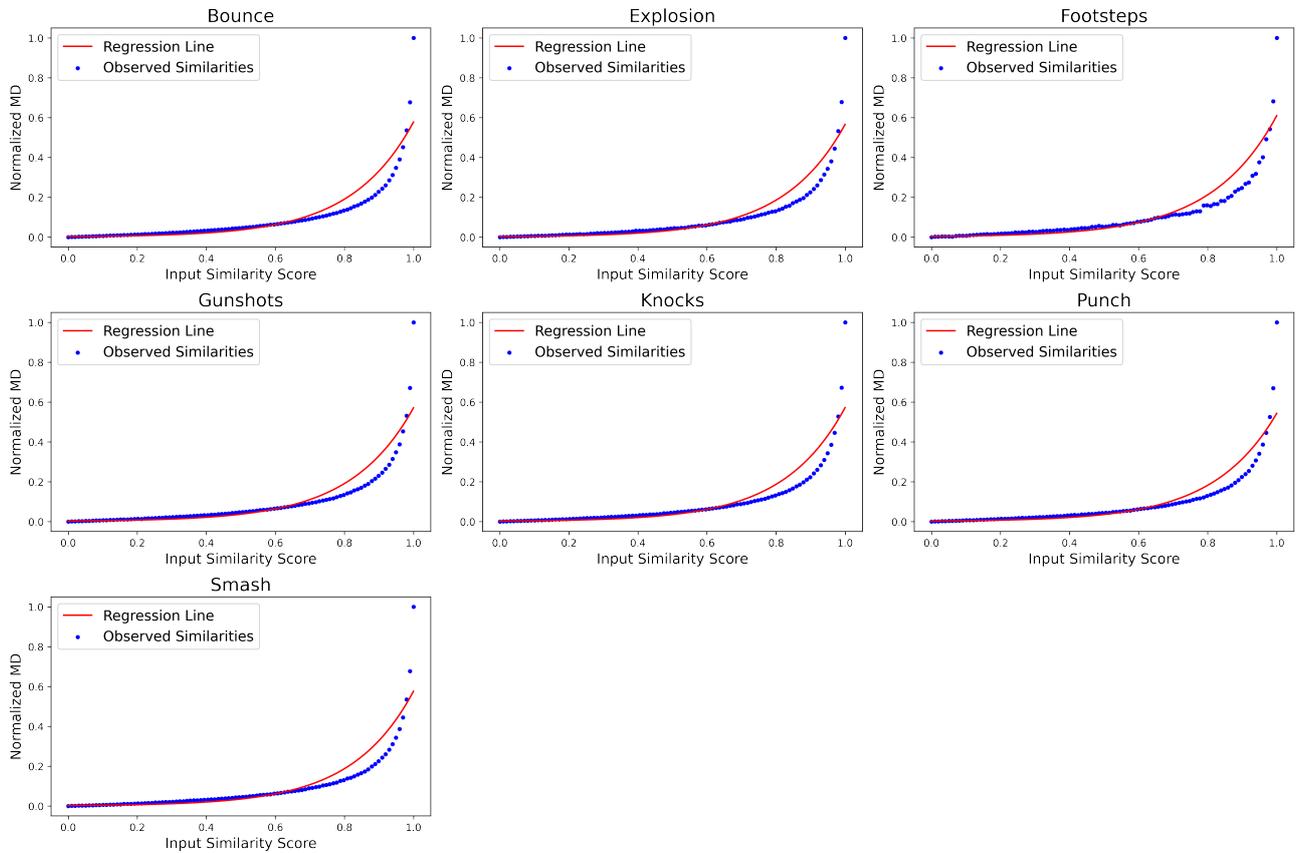

Fig. 9. Regression Analysis for our model trained on Impact-set.

## THE AUTHORS

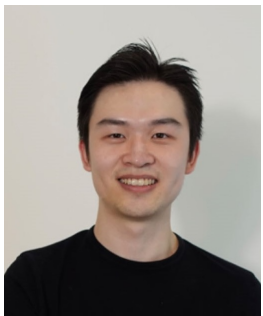

Yunyi Liu

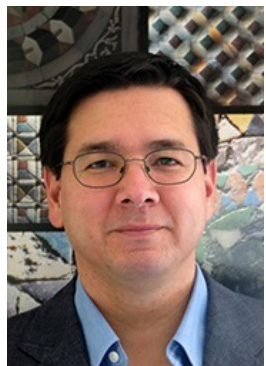

Craig Jin

Yunyi Liu is a Phd student at University of Sydney in Electrical and Information Engineering school, Sydney, Australia. He received a bachelor of music and sound design degree in University of Technology Sydney and a master of interaction design and electronic arts degree in University of Sydney in 2019 and 2021 respectively. His Phd topic is 'interpretable and controllable sound effects generation with neural audio synthesis'. He has been researching on conditioning methods for audio generative models with a focus of unsupervised learning by extracting meaningful information from limited datasets. His research interests include neural audio synthesis, controllable and interpretable generative models, audio signal processing, machine learning, self-supervised learning, and cross-modal sound syntheis.

Craig Jin received the M.S. degree in applied physics from Caltech, Pasadena, California, in 1991 and the Ph.D. in electrical engineering from the University of Sydney, Sydney, Australia, in 2001. He is an Associate Professor in the School of Electrical and Computer Engineering, University of Sydney and Director of the Computing and Audio Research Laboratory. His research interest includes experimental and theoretical aspects of acoustic and biomedical signal processing, with a recent focus on context-aware computing, assistive technologies and auditory sensory augmentation. He has authored more than 240 papers in these research areas, holds eight patents, and founded three start-up companies. He received national recognition in Australia (April 2005, Science in Public Fresh Innovators Program) for his invention of a spatial hearing aid.